\documentclass[a4paper]{jpconf}
\usepackage{graphicx}
\begin{document}
\title{Towards new frontiers in CP violation in B decays}

\author{Robert Fleischer}

\address{Nikhef, Science Park 105, NL-1098 XG Amsterdam, Netherlands}

\address{Department of Physics and Astronomy, Vrije Universiteit Amsterdam, 
NL-1081 HV Amsterdam, Netherlands}

\ead{Robert.Fleischer@nikhef.nl}

\begin{abstract}
CP-violating effects in decays of $B$ mesons offer a wide spectrum of probes for testing the 
phase structure of the quark-flavour sector of the Standard Model. After a brief discussion of 
the picture emerging from the current LHC data, the focus will be put on two specific topics: 
hadronic uncertainties from penguin topologies on measurements of the $B^0_q$--$\bar B^0_q$ 
mixing phases ($q\in\{d,s\}$), and the $U$-spin-related decays $B^0_s\to K^+K^-$ 
and $B^0_d\to\pi^+\pi^-$. Valuable new insights are expected from future studies of 
CP violation in $B$ decays. For the detection of possible new sources of CP violation, 
it will be crucial to match the experimental and theoretical precisions and to have a careful look
at the underlying assumptions.
\end{abstract}
\section{Introduction}
In the Standard Model (SM), the rich phenomenology of quark-flavour physics and CP 
violation is related to the Cabibbo--Kobayashi--Maskawa (CKM) quark-mixing matrix 
\cite{cab,KM}. Information on the phase structure and elements of this matrix are
encoded in weak decays of $K$, $D$ and $B$ mesons. Since the theory is 
formulated in terms of quarks while the mesons are bound states of strong interactions, 
we have to deal with process-dependent, non-perturbative ``hadronic" parameters in
the calculation of the relevant transition amplitudes. This feature gives rise to 
the main challenge in studies of CP violation in $B$ decays: hadronic uncertainties.

In the presence of New Physics (NP), typically new sources of flavour and CP violation 
arise. Analyses of weak meson decays are facing an impressive hierarchy of scales:
\begin{equation}
\underbrace{\Lambda_{\rm NP}\sim 10^{(0 ... ?)} \,\mbox{TeV} \,  \gg \,
\Lambda_{\rm EW}\sim 10^{-1} \,\mbox{TeV}}_{\mbox{short distance physics}}
\quad \gg\gg \, \underbrace{\Lambda_{\rm QCD}\sim 10^{-4} 
\,\mbox{TeV}}_{\mbox{long distance physics}}.
\end{equation}
In order to deal with this situation, effective field theories offer the suitable theoretical tool. 
Within this framework, the heavy degrees of freedom (NP particles,  top quark, $Z$ and $W$ bosons)
are integrated out from appearing explicitly and are described in short-distance loop functions. 
Perturbative QCD corrections can be calculated in a systematic way, and renormalisation group
techniques allow the summation of large logarithms. This machinery was applied to the SM and 
various popular NP scenarios, such as MSSM, models with universal and warped extra 
dimensions, little Higgs models, scenarios with extra $Z'$ bosons, etc., as reviewed in
\cite{BG-rev}.

Following these lines, low-energy effective Hamiltonians can be calculated for $\bar B \to \bar f$ 
processes, taking the following general form \cite{BBL}:
\begin{equation}
\langle \bar f|{\cal H}_{\rm eff}|\bar B\rangle=
\frac{G_{\rm F}}{\sqrt{2}}\sum_{j}\lambda_{\rm CKM}^{j}\sum_{k}
C_{k}(\mu)\,\langle \bar f|Q^{j}_{k}(\mu)|\bar B\rangle,
\end{equation}
where $G_{\rm F}$ is Fermi's constant and the $\lambda_{\rm CKM}^{j}$ denote 
combinations of CKM matrix elements. The short-distance contribution to the decay 
amplitude is described by the 
Wilson coefficient functions $C_{k}(\mu)$, which can be calculated in perturbation theory for 
the SM and its extensions. On the other hand, the hadronic matrix elements 
$\langle \bar f|Q^{j}_{k}(\mu)|\bar B\rangle$ are non-perturbative quantities, describing the
long-distance contributions.

The key players for the exploration of CP violation in $B$ decays are non-leptonic channels. 
In the previous decade, there were interesting developments for calculations of 
such processes within QCD: QCD factorisation, the perturbative hard-scattering (PQCD)
approach, soft collinear effective theory (SCET) and QCD sum rules; the state of the art is
discussed in \cite{Bell}. There has recently been impressive progress in lattice QCD \cite{Juettner}. 
However, non-leptonic $B$ decays generally remain a theoretical challenge, 
which is also indicated by experimental data.
 
 The outstanding feature of analyses of CP violation in $B$ decays is that the calculation of
 the hadronic matrix elements $\langle \bar f|Q^{j}_{k}(\mu)|\bar B\rangle$
 can be circumvented in fortunate cases (for a detailed discussion, see \cite{RF-rev}).
 The corresponding strategies play a key role to ``over constrain" the  unitarity triangle (UT) 
 of the CKM matrix. Detailed analyses and continuous updates are performed by the 
 CKMfitter \cite{CKMfitter} and UTfit \cite{UTfit} collaborations. The current picture of the UT 
 shows impressive consistency with the CKM sector of the SM, despite a few tensions.
  
The previous run of the Large Hadron Collider (LHC) has resulted in the exciting 
discovery of the Higgs boson. On the other hand, the ATLAS and CMS experiments have 
so far not seen signals of NP particles, and the SM flavour sector has been confirmed 
by the LHCb data, apart from a few discrepancies which are unfortunately not 
yet conclusive. The implications for the general structure of physics beyond the SM
are a large characteristic NP scale, i.e.\ not just $\sim \mbox{TeV}$, or (and?) symmetries 
preventing large NP effects in the flavour sector, where models with ``Minimal Flavour Violation"  
are the most prominent example. 

Many more interesting results are expected from the next run of the 
LHC and its upgrade as well as high-precision flavour experiments. 
In view of the present situation, we have to prepare ourselves to deal with 
smallish NP effects. It will be crucial for resolving possible signals of NP in the data 
to have a careful look at the underlying theoretical assumptions and approximations.
The challenge will be to match the experimental and theoretical precisions.

\section{Penguin effects in benchmark probes of CP violation}
Neutral $B^0_q$ mesons ($q\in\{d,s\}$) show $B^0_q$--$\bar B^0_q$ mixing \cite{Bobeth}, 
which originates from box topologies in the SM but may well receive NP contributions. The 
CP-violating mixing phases are given by
\begin{equation}
\phi_d=\phi_d^{\rm SM}+\phi_d^{\rm NP} = 2\beta + \phi_d^{\rm NP}, \quad
\phi_s=\phi_s^{\rm SM}+\phi_s^{\rm NP} = -2\lambda^2\eta + \phi_s^{\rm NP},
\end{equation}
where $\beta$ is the usual angle of the UT, while $\lambda\equiv|V_{us}|\sim 0.22$ and 
$\eta$ are parameters of the Wolfenstein parametrisation of the CKM matrix. The benchmark
decays to measure the mixing phases $\phi_q$ through mixing-induced CP violation
are given by $B^0_d\to J/\psi K_{\rm S}$, 
$B^0_s\to J/\psi\phi$ and $B^0_s\to J/\psi f_0(980)$. Decays of $B_s$ mesons play 
the key role at the LHC \cite{BFS}.
 
These determinations are affected by uncertainties from doubly Cabibbo-suppressed 
penguin contributions 
[12--19], which cannot be calculated reliably and are usually neglected. 
In view of the current situation and the increasing experimental precision, 
the following questions arise: how important are the penguin contributions and 
how can they be controlled?

\subsection{The $B^0_{d,s}\to J/\psi K_{\rm S}$ system}\label{BJpsiK}
In the SM, the decay $B^0_d\to J/\psi K_{\rm S}$ originates from a colour-suppressed tree
topology and penguin topologies with up, charm and top quarks running in the loops. Using the
unitarity of the CKM matrix, the decay amplitude can be written as follows \cite{RF-psiK}:
\begin{equation}\label{BdpsiK-ampl}
A(B_d^0\to J/\psi\, K_{\rm S})=\left(1-\lambda^2/2\right){\cal A'}
\left[1+\epsilon a'e^{i\theta'}e^{i\gamma}\right],
\end{equation}
where ${\cal A}'$  and $a' e^{i\theta'}$ are CP-conserving parameters, involving the relevant
hadronic matrix elements. Whereas the former quantity is governed by the colour-suppressed 
tree contribution, the latter measures the ratio of penguin to tree topologies.  The key 
feature of (\ref{BdpsiK-ampl}) is that the penguin parameter 
$a'$ enters with the tiny $\epsilon\equiv\lambda^2/(1-\lambda^2)=0.05$; 
$\gamma$ is the usual UT angle.

CP violation is probed through the time-dependent decay rate asymmetry 
\begin{equation}\label{CP-asym}
\frac{\Gamma(B^0_d(t)\to J/\psi K_{\rm S})-
\Gamma(\bar B^0_d(t)\to J/\psi K_{\rm S})}{\Gamma(B^0_d(t)\to J/\psi K_{\rm S})+
\Gamma(\bar B^0_d(t)\to J/\psi K_{\rm S})}
=C_{J/\psi K_{\rm S}}\cos(\Delta M_d t)-S_{J/\psi K_{\rm S}}\sin(\Delta M_d t),
\end{equation}
where the direct CP asymmetry $C_{J/\psi K_{\rm S}}$ is proportional to 
$\epsilon\,a'\sin\theta'\sin\gamma$. On the 
other hand, the mixing-induced CP asymmetry can be written in the following form \cite{FFJM}:
\begin{equation}
 S(B_d\to J/\psi K_{\rm S})=\sin(\phi_d+\Delta\phi_d),
\end{equation}
where the hadronic phase shift $\Delta\phi_d$ is proportional to 
$\epsilon\,a' \cos\theta' \sin\gamma$ (and is usually neglected). 

The decay $B^0_s\to J/\psi K_{\rm S}$ is related to $B^0_d\to J/\psi K_{\rm S}$ through the
$U$-spin symmetry of strong interactions \cite{RF-psiK}. In the SM, its decay amplitude can
be written as
\begin{equation}
A(B_s^0\to J/\psi\, K_{\rm S})=-\lambda\,{\cal A}\left[1- a e^{i\theta} e^{i\gamma}\right].
\end{equation}
In contrast to (\ref{BdpsiK-ampl}), $a$ does not
enter with $\epsilon$, i.e.\ is not doubly Cabibbo-suppressed. Consequently, the penguin effects
are magnified in $B^0_s\to J/\psi K_{\rm S}$. It is useful to introduce a quantity
$H\propto {\rm BR} (B_s \to J/\psi K_{\rm S})/{\rm BR}(B_d\to J/\psi K_{\rm S})$, 
which complements the direct and mixing-induced CP asymmetries of the time-dependent 
CP-violating rate asymmetry of $B^0_s\to J/\psi K_{\rm S}$. 

The $U$-spin symmetry implies $a=a'$ and $\theta=\theta'$, thereby 
allowing the determination of $\gamma$, $a$ and $\theta$ from $H$ and the two CP-violating
observables of $B^0_s\to J/\psi K_{\rm S}$ \cite{RF-psiK}. Since 1999, when this strategy was
originally proposed, there has been a change of the main
focus: a study \cite{DeBFK} has shown that the extraction of 
$\gamma$ will be feasible at LHCb but not competitive with other methods.
As $\gamma$ will be know by the time CP violation in $B^0_s\to J/\psi K_{\rm S}$
can be detected, the corresponding CP asymmetries allow a clean determination 
of the penguin parameters $a$ and $\theta$. 

The $B^0_s\to J/\psi K_{\rm S}$ channel was observed by CDF \cite{CDF-BspsiKS}
and LHCb \cite{LHCb-BspsiKS-BR} but its CP asymmetries have not yet been measured.
Using currently available data for decays with a CKM structure similar to 
$B^0_s\to J/\psi K_{\rm S}$, i.e.\ $B^0_d\to J/\psi \pi^0$
and $B^+\to J/\psi \pi^+$,  and complementing them with $B^0_d\to J/\psi K^0$, 
$B^+\to J/\psi K^+$ data, the size of the penguin parameters $a$ and $\theta$ can be 
constrained. An analysis along these lines yields the following preliminary results \cite{DeBF-pen}:
\begin{equation}\label{pengs}
a=0.17^{+0.13}_{-0.11}, \quad \theta=(182.4^{+4.7}_{-4.6})^\circ, \quad
\Delta\phi_d=(-0.97^{+0.72}_{-0.65})^\circ.
\end{equation}

\subsection{CP violation in $B^0_s\to J/\psi \phi$}
The CKM structure of the $B^0_s\to J/\psi \phi$ channel is analogous to that of 
$B^0_d\to J/\psi K_{\rm S}$. However, as the final state is a mixture of CP-even and CP-odd 
eigenstates $f\in\{0,\parallel,\perp\}$, a time-dependent angular analysis of the 
$J/\psi [\to\mu^+\mu^-]\phi [\to\ K^+K^-]$ decay products has to be performed 
[23--25]. 
The impact of the SM penguin contributions is usually neglected. As in the case of 
$B^0_d\to J/\psi K_{\rm S}$, the expressions for the mixing-induced CP 
asymmetries are modified as follows \cite{FFM}:
\begin{equation}
{\cal A}_{{\rm CP}, f}^{\rm mix}=
\sin\phi_s \,\to\,\sin(\phi_s+\Delta\phi_s^f),
\end{equation}
where the hadronic phase shift $\Delta\phi_s^f$ depends on the final-state configuration $f$.
The current average (neglecting the penguin effects) of the CDF, D0, ATLAS and LHCb data 
compiled by the Heavy Flavour Averaging Group  
is given by $\phi_s=(0.0 \pm 4.0)^\circ$ \cite{HFAG}, 
which agrees with the SM value $\phi_s^{\rm SM}=-(2.11\pm0.08)^\circ$ \cite{CKMfitter}
of the $B^0_s$--$\bar B^0_s$ mixing phase.

A tool to control the penguin effects is offered by $B^0_s\to J/\psi \bar K^{*0}$ \cite{FFM}, 
which was observed by CDF \cite{CDF-BspsiKS} and LHCb \cite{LHCb-BpsiKast}.
Its branching ratio $(4.4^{+0.5}_{-0.4}\pm0.8) \times 10^{-5}$ is found 
in agreement with the prediction $(4.6\pm0.4) \times 10^{-5}$ following from 
$B^0_d\to J/\psi \rho^0$, and its polarisation fractions agree well with those of 
$B^0_d\to J/\psi K^{*0}$. The $B^0_d\to J/\psi \rho^0$ channel, which shows also 
mixing-induced CP violation, is another interesting decay to shed light on the 
hadronic penguin effects \cite{FFM}. 

The experimental sensitivity from $B^0_s\to J/\psi \phi$ at the LHCb upgrade 
(50\,$\mbox{fb}^{-1}$) is expected as 
$\Delta\phi_s|_{\rm exp}\sim 0.008 = 0.46 ^\circ$ \cite{LHCb-strategy}. 
This impressive precision will make it mandatory to get a handle on the penguin effects,
which may lead to phase shifts at the $1^\circ$ level (as indicated by (\ref{pengs})).

\subsection{CP violation in $B^0_s\to J/\psi f_0(980)$}
Another interesting probe to study CP violation is provided by 
$B^0_s\to J/\psi f_0(980)$ \cite{SZ}. In contrast to $B^0_s\to J/\psi \phi$, 
as the $f_0(980)$ is a scalar state with quantum numbers $J^{PC}=0^{++}$,
the final state is present in a $p$ wave and has the CP eigenvalue $-1$. Consequently, 
a time-dependent angular analysis is not needed. On the other hand, the 
hadronic structure of the $f_0(980)$ is still -- after decades -- not settled, with a variety of 
theoretical interpretations ranging from the quark--antiquark picture to tetraquarks. 
A detailed discussion of the implications of this feature for the extraction of $\phi_s$ was given in 
\cite{FKR-I} (for $B^0_{s,d} \to J/\psi \eta^{(\prime)}$ decays, see \cite{FKR-II}), while
recent LHCb measurements related to this topic are reported in \cite{LHCb-recent}.

\subsection{Effective $B^0_s$ decay lifetimes}
The measurement of effective lifetimes of $B^0_s\to f$ decays, which are defined as
\begin{equation}
 \tau_{f} 
  \equiv \frac{\int^\infty_0 t\ \langle \Gamma(B_s(t)\to f)\rangle\ dt}
  {\int^\infty_0 \langle \Gamma(B_s(t)\to f)\rangle\ dt},
\end{equation}
offers yet another way to obtain insights into CP violation \cite{FK-lifetimes}. Here
it is particularly interesting to compare $B^0_s$ decays into CP-odd final states, such as 
$B^0_s\to J/\psi f_0(980)$, with those into CP-even final states, such as 
$B^0_s\to K^+K^-$ and $B^0_s\to D_s^+D_s^-$. 
The measured effective lifetimes can be converted into contours in the $\phi_s$--$\Delta\Gamma_s$
plane, where $\Delta\Gamma_s$ is the decay width difference of the $B_s$-meson 
system (for an overview of the status of $\Delta\Gamma_s$, see \cite{lenz}).
The lifetime contours are very robust with respect to hadronic uncertainties  \cite{FK-lifetimes}. 
For an update with the most recent LHCb data, see \cite{Gandini}. 
The $B_s$ decay lifetimes result in a picture in agreement with the SM.

\subsection{Comments for the LHCb upgrade era}
In view of hadronic effects, it is important to give measurements of $\phi_s$ for the individual 
decay channels $B^0_s\to f$, i.e.\ $B^0_s\to J/\psi \phi$ and $B^0_s\to J/\psi f_0(980)$. 
The pattern of the $(\phi_s)_f$ may provide insights into the hadronic effects: differences 
in the values of $(\phi_s)_f$ would indicate hadronic effects. On the other hand, should 
no differences between the individual $\phi_s$ emerge, there would be evidence for negligible 
hadronic effects (within the errors) or a universal hadronic phase shift. 

The time-dependent analysis of CP violation in $B^0_s\to J/\psi K_{\rm S}$ allows the 
clean determination of the corresponding penguin parameters (see Subsection~\ref{BJpsiK}). 
A sizeable penguin parameter $a$ would indicate a potential problem in the measurement of 
$\phi_s$ from the $B^0_s\to J/\psi \phi$ and $B^0_s\to J/\psi f_0(980)$ channels. On the 
other hand, smallish penguin effects would give us confidence for the measurement 
of $\phi_s$, although subtleties may arise due to the different final states.

\section{CP Violation in \boldmath$B^0_s\to K^+K^-$\unboldmath\ and 
\boldmath$B^0_d\to\pi^+\pi^-$\unboldmath}
The decays $B^0_s\to K^+K^-$ and $B^0_d\to\pi^+\pi^-$ receive contributions from
tree and penguin topologies. In the SM, their decay amplitudes can be written as
\begin{equation}
A(B_s^0\to K^+K^-)\propto {\cal C}'\left[ e^{i\gamma}+
d'e^{i\theta'}/\epsilon\right], \quad
A(B_d^0\to\pi^+\pi^-)\propto {\cal C}\left[ e^{i\gamma}-d\,e^{i\theta}\right],
\end{equation}
where $d'e^{i\theta'}$, ${\cal C}'$ and their unprimed counterparts
are CP-conserving strong quantities \cite{RF-BsKK}. The direct and
mixing-induced CP asymmetries of the $B^0_s\to K^+K^-$ and $B^0_d\to\pi^+\pi^-$
decays allow the determination of theoretically clean contours in the 
$\gamma$--$d'$ and $\gamma$--$d$ planes, respectively. Since these decays are 
related to each other through the interchange of all down and strange quarks, the 
$U$-spin symmetry of strong interactions implies $d'=d$ and $\theta'=\theta$;
the former relation allows the extraction of $\gamma$ and $d(=d')$ from the contours
\cite{RF-BsKK,RF-BsKK-07}. Moreover, the strong phases $\theta$ and $\theta'$ can be 
determined, allowing an internal consistency check of the $U$-spin symmetry. Further insights
into the hadronisation dynamics are provided by $|{\cal C}|$ and $|{\cal C}'|$, which can
be extracted from the ratio $K\propto \mbox{BR}(B_s\to K^+K^-)/\mbox{BR}(B_d\to\pi^+\pi^-)$.
This strategy is promising for the LHCb physics programme \cite{LHCb-strategy}. 
It will be particularly interesting to compare the resulting value of $\gamma$ with those following 
from methods using only tree-diagram-like $B_{(s)}$-meson decays. 

The picture resulting from the current data was explored in detail in \cite{RF-BsKK-07,FK-Uspin}; 
the numerical results given below refer to the update by Rob Knegjens in \cite{Knegjens-PhD}. 
An interesting variant of the method was proposed in \cite{CFMS}, as discussed in detail by 
Marco Ciuchini in \cite{Ciuchini-Beauty}.

Using information on $K$, CP violation in 
$B^0_d\to\pi^+\pi^-$ and $B^0_d\to\pi^\mp K^\pm$, and allowing for $U$-spin-breaking
corrections 
$ \xi\equiv d'/d = 1 \pm 0.15$,  $\Delta\theta\equiv\theta'-\theta = \pm 20^\circ$ results in
\begin{equation}
\gamma=(67.7^{+4.5}_{-5.0}|_{\rm input}
  \mbox{}^{+5.0}_{-3.7}|_\xi\mbox{}^{+0.1}_{-0.2}|_{\Delta\theta})^\circ,
\end{equation}
which agrees with the ``tree-level" results $\gamma=(70.0^{+7.7}_{-9.0})^\circ$ 
\cite{CKMfitter} and $(69.4 \pm 7.1)^\circ$ \cite{UTfit} within the uncertainties. 
There are no indications for sizeable non-factorisable $SU(3)$-breaking corrections in 
the corresponding data. In the SM, the mixing-induced CP asymmetry is predicted as 
\begin{equation}
{\cal A}_{\rm CP}^{\rm mix}(B_s\to K^+K^-)|_{\rm SM}=\ -0.220^{+0.042}_{-0.054},
\end{equation}
while ${\cal A}_{\rm CP}^{\rm dir}(B_s\to K^+K^-) \approx 
{\cal A}_{\rm CP}^{\rm dir}(B_d\to \pi^\pm K^\pm)=0.082\pm0.04$. 

The first LHCb measurement \cite{LHCb-BsKK} of the CP-violating $B_s\to K^+K^-$  
observables yields 
\begin{equation}
{\cal A}_{\rm CP}^{\rm mix}(B_s\to K^+K^-)=  -0.30 \pm 0.12 \pm 0.04,\quad
{\cal A}_{\rm CP}^{\rm dir}(B_s\to K^+K^-)= 0.14\pm0.11\pm0.03,
\end{equation}
and agrees with the SM predictions given above. 
In the future, once the experimental precision for the CP asymmetries improves, $\gamma$ can 
be extracted exclusively from the $\gamma$--$d^{(')}$ contours. The observable $K$, which  
is affected by form factors and non-factorisable effects, will then yield insights into hadronic physics. 
The current data point towards a fortunate situation for the determination of $\gamma$ which is very robust with respect to $U$-spin-breaking 
corrections  [36--38].  

\section{Outlook}
The exploration of CP violation in $B$ decays is a very broad field, with many other 
interesting topics complementing those discussed above. I would like to briefly give
two more examples:
\begin{itemize}
\item The penguin decay $B^0_s\to\phi\phi$ 
(for a recent theoretical discussion, see \cite{BDDL}). This summer, LHCb 
announced the first time-dependent angular analysis 
of CP violation this channel \cite{Kristof}. The result 
$\phi_s=-0.17 \pm 0.15 ({\rm stat}) \pm 0.03 ({\rm syst}) = - (9.7 \pm  8.8)^\circ$ 
is consistent with the SM at the present 
level of precision. This channel has a lot of potential for the future. 
\item A promising decay for Belle II at SuperKEKB is the decay 
$B^0_d\to \pi^0 K_{\rm S}$. A correlation between the direct and mixing-induced
CP asymmetries of this channel can be predicted in the SM, with current data showing 
an intriguing discrepancy \cite{FJPZ}. It will be interesting to monitor the future measurements 
of the corresponding observables at Belle II.
\end{itemize}

We are moving towards new frontiers in particle physics. There are still no unambiguous 
signals for NP at the LHC, and it is impressive -- also frustrating -- to see how the SM 
stands more and more stringent tests, both at the high-energy and at the high-precision frontier. 
Much more is yet to come with the future running of the LHC and dedicated studies of 
flavour physics, including CP violation in $B$ decays. 
However, we have to prepare ourselves to deal with smallish NP effects
in the data. In view of the increasing experimental precision, we have to be careful 
with respect to theoretical assumptions and approximations. The challenge will be the 
matching of the experimental and theoretical uncertainties in the future high-precision era. 
Interesting and fruitful years for the further testing of the SM and the search of NP are 
ahead of us!

\vspace*{0.5truecm}

\noindent {\it Acknowledgements}\\
I would like to thank Cristina Lazzeroni and her co-organizers for inviting me to this 
most interesting and enjoyable conference and their kind hospitality in Birmingham.

\section*{References}
\end{document}